\documentclass[12pt,tightenlines,eqsecnum,floats,aps,amsmath,amssymb,nofootinbib,prd,superscriptaddress,showpacs]{revtex4}

\usepackage{setspace}
\usepackage{amsmath,amssymb,amsfonts,amsthm}
\usepackage{graphicx}
\usepackage{enumerate}

\newtheorem{thm}{Theorem}[section]
\newtheorem{lem}[thm]{Lemma}
\newtheorem{Def}{Definition}[section]

\makeatletter
\newcommand*{\simboloG}[1]{%
  \vphantom{\sum}
  \smash{%
    \mathchoice{%
      \raisebox{-.3\height}{\Huge$\m@th\displaystyle#1$}%
      }{%
      \raisebox{-.2\height}{\LARGE$\m@th#1$}%
      }{%
      \raisebox{-.2\height}{\LARGE$\m@th#1$}%
      }{%
      \raisebox{-.2\height}{\LARGE$\m@th#1$}%
      }%
    }}
\newcommand{\BigTimes}{\mathop{\simboloG{\times}}}
\makeatother

\begin{document}

\title{On the computation of black hole entropy in loop quantum gravity}

\author{J. Fernando \surname{Barbero G.}}
\email[]{fbarbero@iem.cfmac.csic.es} \affiliation{Instituto de
Estructura de la Materia, CSIC, Serrano 123, 28006 Madrid, Spain}

\author{Eduardo J. \surname{S. Villase\~nor}}
\email[]{ejsanche@math.uc3m.es} \affiliation{Instituto Gregorio Mill\'an, Grupo de Modelizaci\'on
y Simulaci\'on Num\'erica, Universidad Carlos III de Madrid, Avda.
de la Universidad 30, 28911 Legan\'es, Spain} \affiliation{Instituto
de Estructura de la Materia, CSIC, Serrano 123, 28006 Madrid, Spain}

\date{October 9, 2008}

\begin{abstract}
We discuss some issues related to the computation of black hole entropy in loop quantum gravity from the novel point of view provided by the recent number-theoretical methods introduced by the authors and their collaborators. In particular we give exact expressions, in the form of integral transforms, for the black hole entropy in terms of the area. We do this by following several approaches based both on our combinatorial techniques and also on functional equations similar to those employed by Meissner in his pioneering work on this subject. To put our results in perspective we compare them with those of Meissner. We will show how our methods confirm some of his findings, extend the validity of others, and correct some mistakes. At the end of the paper we will discuss the delicate issue of the asymptotics of black hole entropy.
\end{abstract}

\pacs{04.70.Dy, 04.60.Pp, 02.10.Ox, 02.10.De}

\maketitle

\section{Introduction}\label{intro}

The derivation of the Hawking-Bekenstein area law for realistic black holes is one of the most important achievements of loop quantum gravity (LQG) \cite{abk,DL,M}. This, together with the recent insights on the Big Bang singularity provided by loop quantum cosmology \cite{Boj}, are two of the physical pillars on which the formalism is currently supported. The fact that areas are quantized in LQG is another result that plays a relevant role in the study of black hole physics. This is so because, as pointed out in classic papers by Bekenstein and Mukhanov \cite{BM}, there are good reasons to believe that black hole areas should be quantized in such a way that the spacing between consecutive values of them is constant. Here, however, LQG does not seem to provide a completely satisfactory description because the spectrum of the area operator is not equally spaced.

A surprising development in this matter took place when Corichi, D\'{\i}az-Polo, and Fern\'andez-Borja \cite{val1} found, by using a direct computer intensive approach, that \textit{effectively} the black hole degeneracy spectrum for small black holes can be considered to be equally spaced. This intriguing result --obtained for black holes of around a hundred Planck areas-- is beautiful because it suggests that, after all, the expected behaviour for the entropy can be somehow obtained within the LQG framework. However appealing this finding may seem, it should be taken with some care because the description used for black holes in LQG, modeled by isolated horizons (IH), can only be approximate\footnote{Notice, for example, that microscopic black holes should evaporate very quickly; a fact that cannot be easily taken into account with the techniques currently available in LQG.}, though arguably good for sufficiently large objects. It is then very important to find out if the observed microscopic behaviour of the entropy is also present for macroscopic objects. This would be a very nice result because the Beckenstein-Mukhanov prediction would be non-trivially realized within LQG in a macroscopic regime where the model used to describe black holes is arguably accurate. The main obstacle to find out if this is the case is the impossibility of extending the numerical methods used to date to these large scales. Instead, one must rely on a suitable asymptotic analysis based on closed and explicit expressions for the black hole entropy.

This problem has been already considered in the literature; in fact a solution to it is proposed in a well known paper by K. Meissner \cite{M}. According to the result presented there the asymptotic behaviour of the entropy \textit{does not} display the oscillations described in \cite{val1}.
What is happening then?
Is the behaviour observed in \cite{val1} an artifact of the algorithm used to compute the black hole entropy?
Is there an independent way to check it?

An answer for these questions appears in \cite{prlnos} where a new method to compute black hole degeneracies based on a number-theoretical approach is developed. The procedure proposed in that paper provides an algorithm that can be used to check and extend previous numerical results. The conclusion is unambiguous: the results on the entropy originally found in \cite{val1} are quantitatively correct and persist up to areas an order of magnitude larger than the ones considered in that paper\footnote{Actually up to the largest areas that we have been able to reach by using laptops or personal computers.}.  The structure in the entropy spectrum is clearly present in the new regime explored in \cite{prlnos} and the constancy of the effective spacing between the areas is confirmed.

Once the reality of this effect has been settled beyond doubt it is natural to pose an additional set of  questions, for example: is it possible to find a single procedure that allows us to derive both the microscopic results of \cite{val1,prlnos} and the expressions given by Meissner for the macroscopic black hole entropy?
Can we trust the asymptotic analysis performed by Meissner that seems to exclude, in the large area regime, the behaviour found for small black holes?

Some steps towards answering the first of these questions have been taken in \cite{EF} where the new number-theoretical methods introduced in \cite{prlnos} were used to obtain generating functions for the black hole degeneracy spectrum. This is an important starting point because once closed expressions for them are available it is possible to obtain exact expressions for the black hole entropy in terms of the area. The first goal of this paper is to explain how this can be done.

After the \textit{exact} expression of the black hole entropy is obtained we will compare it with the ones given in \cite{M}. As we will see, when the projection constraint is not taken into account, we exactly reproduce\footnote{In order to be fully consistent with the definition of entropy that we are using here a term $s^{-1}$ should be added to formula (13) of \cite{M}. We will comment on this later.} the result given by formula (13) of \cite{M}. On the other hand when the projection constraint is incorporated our results disagree with those of Meissner. In order to understand the source of this discrepancy we solve the problem from scratch by using functional equations as done in \cite{M}. This is the second goal of the paper, namely, rederive the results of Meissner by using the right functional equations. As we show there is an error in the relation used by him when the projection constraint is included. We identify the source of this error and correct it. Furthermore, the inclusion of a superfluous prefactor in the equation given in \cite{M} makes it difficult to solve in closed form. As we will see it is possible to write a simpler --but equivalent-- functional equation and solve it \textit{exactly} in a straightforward way. After doing this we get the same expression for the black hole entropy that we find by using our generating functions. In our opinion this provides a solid confirmation of our results and highlights the power of the combinatorial and number-theoretical methods of \cite{prlnos,EF}. Along the way we show that, at variance with the claims of Meissner\footnote{According to Meissner his formula (35) is obtained after neglecting some contributions in the resolution of the functional equation that gives the entropy.}, it is in fact possible to give exact expressions for the entropy, without using any approximation, also when the projection constraint is taken into account.

After finding exact expressions for the black hole entropy as suitable integral transforms, we want to raise some points concerning their asymptotic behaviour and then answer the second question posed before. We do this only in the case where the projection constraint is not used (the difficult problem of obtaining the asymptotic behaviour of the entropy in full generality will be addressed elsewhere). Notice, however, that as shown in \cite{val1,val2}, the interesting structure of the black hole entropy is present even if the projection constraint is not taken into account so it makes sense to concentrate on this somewhat simpler situation. The most important issue to discuss now has to do with the poles of the Laplace transform of the entropy. The main result that we prove in the paper in this regard is that the \textit{real parts} of these poles have an accumulation point precisely for the value of the Immirzi parameter $\tilde{\gamma}_M$ computed by Meissner. In order to show this we need some auxiliary results concerning the distribution of poles that we will explicitly write down and prove. The fact that the real parts accumulate to $\tilde{\gamma}_M$ makes the asymptotic study of the entropy highly non-trivial as we will discuss. The bottom line is that the behaviour of the entropy in terms of the horizon area $A$ is somehow proportional to $\exp(\tilde{\gamma}_M A)$, however it is not possible to exclude the possibility that this behaviour is modulated by an oscillatory, possibly decaying, term. Understanding the details of this behaviour
is a crucial issue in order to see if the structure found in the entropy spectrum for small black holes is present in the macroscopic regime. We want to stress at this point that, according to our results, there are good reasons to believe that the value of the Immirzi parameter $\tilde{\gamma}_M$ computed by Meissner is actually the correct one. At any rate, a final statement on this fact can only be made when the full asymptotic behaviour of the entropy is found.

The paper is organized as follows: After this introduction we start with section \ref{entropy} where we discuss in detail the obtention of black hole entropy in LQG. The presentation that we give here is complementary to the one appearing in \cite{prlnos} where we gave a unified treatment for the different types of countings proposed in the literature. Here we will use only the standard entropy definition of black hole entropy in LQG as spelled out in \cite{DL}. We show in section \ref{From} how one can obtain exact formulas for the black hole entropy by using the generating functions appearing in \cite{EF}. Section \ref{functional} considers the same problem by using functional equations in the spirit of Meissner. As we will show we recover the results obtained in section \ref{entropy} without resorting to any approximation. These are then compared to the ones obtained by Meissner in \cite{M}. Section \ref{poles} is devoted to a study of the inverse Laplace transform giving the black hole entropy in the simplified case where no projection constraint is used. In particular we enunciate and prove several lemmas concerning the distribution of the poles of the integrand. We will prove an important result on this: the real parts of the poles have an accumulation point precisely at $\tilde{\gamma}_M$. This may invalidate the conclusion that the asymptotics of the entropy as a function of the area is proportional to $\exp(\tilde{\gamma}_M A)$ as will be shown with a concrete example. We end the paper in section \ref{conclusions} with our conclusions, comments, and a short review of the problems that remain to be solved in order to fully understand the macroscopic behaviour of black hole entropy within LQG.

\section{Computation of the entropy in LQG: the black hole degeneracy spectrum}\label{entropy}

This section describes the number-theoretical and combinatorial approaches of \cite{prlnos,EF} to compute black hole entropy in LQG. Here we will consider only the standard counting of \cite{DL}. Quoting almost \textit{verbatim} from this paper we take the following definition:

\bigskip

\begin{Def}
The entropy $S(a)$ of a quantum horizon of the classical area a, according to Quantum Geometry
and the Ashtekar-Baez-Corichi-Krasnov  framework \cite{abk}, is
$$S(a) = \log \mathfrak{n}(a)\,,$$
where $\mathfrak{n}(a)$ is 1 plus the number of all the finite, arbitrarily long, sequences $\vec{m}=(m_1,\ldots,m_n)$ of non-zero half integers, such that the following equality and inequality are satisfied:
$$\sum_{i=1}^n m_i=0, \quad \sum_{i=1}^n\sqrt{|m_i|(|m_i|+1)}\leq \frac{a}{8\pi\gamma\ell_P^2}.$$
Here $\gamma$ is the Immirzi parameter of Quantum Geometry and $\ell_P$ the Planck length. The extra term 1 above comes from the trivial sequence.
\end{Def}

\bigskip

Let us start by introducing some notations, unit conventions, and definitions. In the following
\begin{eqnarray*}
\mathbb{N}=\{1,2,3,\dots\}\,,\quad \mathbb{N}_0=\{0,1,2,3,\dots\}\,,\quad \mathbb{Z}=\{0,\pm 1,\pm 2,\cdots\}\,,\quad \mathbb{Z}_*=\{\pm 1,\pm 2,\cdots\}\,.
\end{eqnarray*}
We will also define
$$
\mathbb{Z}/2:=\{0,\pm 1/2,\pm 1,\pm 3/2,\cdots\}
$$
with analogous definitions for $\mathbb{N}/2$, $\mathbb{N}_0/2$, and $\mathbb{Z}_*/2$. The Kronecker symbol is written as $\delta(i,j)$ and $\theta(x):={\displaystyle{\chi}}_{[0,\infty)}(x)$ denotes the step function that we use throughout the paper (the characteristic function of $[0,\infty)$ satisfying $\theta(0)=1$). Finally $\lfloor x\rfloor$  denotes the integer part (floor) of the real number $x$. In our previous work on this subject \cite{prlnos,EF} we have used units such that $4\pi\gamma \ell^2_P=1$. Here, however, we will take $8\pi\gamma \ell^2_P=1$ to facilitate the direct comparison of our results with those of Meissner. A simple translation guide between expressions written in the two different unit systems can be given: in order to take formulas from the $8\pi\gamma \ell^2_P=1$ unit system to the $4\pi\gamma \ell^2_P=1$ one, it is enough to substitute the areas $a$ appearing in them by $a/2$.

It is convenient now to define several sets that will play a relevant role in the following. First, given  $(a,p)\in [0,\infty)\times \mathbb{Z}/2$, let $\mathcal{N}_\leq(a,p)$ and $\mathcal{N}_\leq(a)$ be the sets
\begin{eqnarray*}
\mathcal{N}_\leq(a,p)&:=&\{\vec{m}\,|\,\exists n\in \mathbb{N}\,:\,  \vec{m}\in (\mathbb{Z}_*/2)^n\,,\, \sum_{i=1}^n\sqrt{|m_i|(|m_i|+1)}\leq a\,,\, \sum_{i=1}^n m_i=p\} \,,\\
\mathcal{N}_\leq(a)&:=&\{\vec{m}\,|\,\exists n\in \mathbb{N}\,:\,  \vec{m}\in (\mathbb{Z}_*/2)^n\,,\, \sum_{i=1}^n\sqrt{|m_i|(|m_i|+1)}\leq a\}\,,
\end{eqnarray*}
and lets us denote by $N_\leq(a,p)$ and $N_\leq(a)$ their respective cardinalities.
Notice that the entropy $S(a)$ is given by\footnote{The number 1 introduced in this definition was not taken into account in \cite{M} so the formulas appearing in that paper should be corrected accordingly.}
\begin{equation}
e^{S(a)}=\mathfrak{n}(a)=N_\leq(a,0)+1\,.
\label{entropylog}
\end{equation}
The set $\mathcal{N}_\leq(a)$ can be written as the disjoint union
$$
\mathcal{N}_\leq(a)=\bigcup_{p\in \mathbb{Z}/2}\mathcal{N}_\leq(a,p)
$$
and hence the cardinality $N_\leq(a)$ can be obtained in terms of the numbers $N_\leq(a,p)$ by summing in $p$, i.e.
$$
N_\leq(a)=\sum_{p\in \mathbb{Z}/2}N_\leq(a,p)\,.
$$

We will also consider the sets
\begin{eqnarray*}
\mathcal{N}(a,p)&:=&\{\vec{m}\,|\,\exists n\in \mathbb{N}\,:\,  \vec{m}\in (\mathbb{Z}_*/2)^n\,,\, \sum_{i=1}^n\sqrt{|m_i|(|m_i|+1)}= a\,,\, \sum_{i=1}^n m_i=p\}\,,\\
\mathcal{N}(a)&:=&\{\vec{m}\,|\,\exists n\in \mathbb{N}\,:\,  \vec{m}\in (\mathbb{Z}_*/2)^n\,,\, \sum_{i=1}^n\sqrt{|m_i|(|m_i|+1)}= a\}=\bigcup_{p\in \mathbb{Z}/2}\mathcal{N}(a,p)\,,
\end{eqnarray*}
whose cardinalities $N(a,p):=|\mathcal{N}(a,p)|$ and $N(a):=|\mathcal{N}(a)|$ satisfy
$$
N(a)=\sum_{p\in \mathbb{Z}/2}N(a,p)\,.
$$
It is clear that the $\mathcal{N}_\leq$-sets can be written as disjoint unions of $\mathcal{N}$-sets. Explicitly
$$
\mathcal{N}_\leq(a,p)=\bigcup_{a'\leq a} \mathcal{N}(a',p)\,,\quad \mathcal{N}_\leq(a)=\bigcup_{a'\leq a} \mathcal{N}(a')=\bigcup_{a'\leq a}\bigcup_{p\in \mathbb{Z}/2} \mathcal{N}(a',p)\,,
$$
and hence
\begin{eqnarray*}
N_\leq(a,p)=\sum_{a'\leq a} N(a',p)\,,\quad N_\leq(a)=\sum_{a'\leq a} N(a')=\sum_{a'\leq a}\sum_{p\in \mathbb{Z}/2} N(a',p)\,.
\end{eqnarray*}

Notice that in order to compute the black hole entropy according to the definition given above we only need to know $N_{\leq}(a,0)$. However, it is convenient at times to work with the $N_{\leq}(a,p)$ so we will keep the $p$-label in the following and impose the condition $p=0$ only when needed.

We will next obtain \textit{exact} formulas for $N(a)$, $N(a,p)$, $N_\leq(a)$, and $N_\leq(a,p)$. The first two, $N(a)$ and $N(a,p)$, refer to what we call the \textit{black hole degeneracy spectrum} \cite{val2} whereas the last two are directly related to the black hole entropy. Before giving a formal derivation we will summarize the procedure that we will use. The first step in all the cases is determining the finite sequences of arbitrary length $n$ consisting of non-zero, positive, half integers $|m_i|$, $i=1,\ldots,n$, satisfying the condition
$$
\sum_{i=1}^n\sqrt{|m_i|(|m_i|+1)}= a
$$
for a given value of $a$. This can be done by first finding out the possible values for the $|m_i|$ (with their multiplicities) compatible with $a$ and then considering all the distinguishable permutations of them. This first step basically solves the problem of obtaining $N(a)$. To obtain $N(a,p)$ one must take into account the \textit{projection constraint} $\sum_{i=1}^nm_i=p$. As explained in \cite{EF} this can be easily done by using generating functions in the form of Laurent polynomials. Finally, to find $N_\leq(a)$ and $N_\leq(a,p)$ one has to give a method to add the cardinalities given by $N(a^\prime)$ and $N(a^\prime,p)$ for all the eigenvalues $a^\prime$ in the area spectrum smaller or equal to the given $a$.

Let us start by considering the sets $\mathcal{N}(a)$ and $\mathcal{N}(a,p)$. We want to give generating functions for both $N(a,p)$ and $N(a)$. To this end, let us consider first the auxiliary set
\begin{eqnarray*}
\mathcal{K}(a)&:=&\{\vec{k}\,|\,\exists n\in \mathbb{N}\,:\,  \vec{k}\in \mathbb{N}^n\,,\, \sum_{i=1}^n\sqrt{(k_i+1)^2-1}= 2a\,\}\,.
\end{eqnarray*}
Now, there exists a surjective map  $\pi:\mathcal{N}(a)\rightarrow \mathcal{K}(a)$ defined by
$$
\mathcal{N}(a)\cap (\mathbb{Z}_*/2)^n\ni \vec{m}\mapsto  \pi(\vec{m})=\vec{k}\in\mathcal{K}(a)\cap \mathbb{N}^n\,,\, \textrm{ where } k_i:=2|m_i|\,,\quad i=1,\dots, n.
$$
Clearly,  if $\vec{m}\in \mathcal{N}(a)$ and $\vec{k}=\pi(\vec{m})$ we get
$$
\sum_{i}\sqrt{|m_i|(|m_i|+1)}= a\Rightarrow \sum_{i}\sqrt{(k_i+1)^2-1}=2a\,.
$$
The map $\pi$ is not injective and hence given  $\vec{k} \in\mathcal{K}(a)\cap \mathbb{N}^n$ it is not possible to unambiguously reconstruct $\vec{m}$ because there are several $\vec{m}\in \mathcal{N}(a)$ such that $\pi(\vec{m})=\vec{k}$ (i.e. several acceptable choices for the signs of the $m_i$),
$$
\vec{m}\in \pi^{-1}(\vec{k})\Leftrightarrow m_i\in\{-k_i/2,k_i/2\}\,,\quad i=1,\dots, n.
$$
In order to determine the cardinality $K(a)=|\mathcal{K}(a)|$ let us fix $a\in[0,\infty)$ and consider the equation
\begin{equation}
\sum_{i=1}^n\sqrt{(k_i+1)^2-1}=\sum_{k=1}^{k_{\mathrm{max}}} n_k\sqrt{(k+1)^2-1}=2a\,,
\label{e1}
\end{equation}
where $n_k$ denotes the number of times that the integer $k\in \mathbb{N}$ appears in the sequence $\vec{k}$. Equation (\ref{e1}) should be understood as an equation in the set of unknowns $\{(k,n_k)\}$. It is important to realize that once the set of all possible solutions is determined the sequences $\vec{k}$ can be found by considering all the permutations  of a multiset where each $k$ appears $n_k$ times.

Notice that we can always write $\sqrt{(k+1)^2-1}$ as the product of an integer times the square root of a square-free positive integer number (SRSFN) by using its prime factor decomposition. Hence, $K(a)>0$ implies that $a$ is constrained to have the form
\begin{equation}
2a=\sum_{i=1}^r q_i\sqrt{p_i}
\label{areaequation}
\end{equation}
where $q_i\in \mathbb{N}_0$ and $p_i>1$ are square-free integers (we will consider that $p_1=2<p_2=3<\cdots <p_i<p_{i+1}<\cdots$, etc). In order to proceed from here we must first identify the allowed values of $k$ such that $\sqrt{(k+1)^2-1}$ is an integer multiple of some $\sqrt{p_i}$, and then determine the value of $n_k$ that tells us how many times each of them appears. We deal with the first  problem by solving the Pell equations associated to each of the SRSFN's in the r.h.s. of (\ref{areaequation}), i.e.
\begin{equation}
\label{pell} \sqrt{(k+1)^2-1}=y \sqrt{p_i}\,\Leftrightarrow (k+1)^2-p_iy^2=1,
\end{equation}
with $y\in\mathbb{N}$. The solutions can be labeled as $\{(k_\alpha^i,y_\alpha^i)\,:\,\alpha\in\mathbb{N}\}$, where here the index $i$ refers to the square-free numbers in each of the Pell equations (see, for instance, \cite{Burton} for details on the Pell equation).  Once these numbers are known the $n_k$ can be found \cite{prlnos} by solving the system of $r$-uncoupled, linear, diophantine equations
\begin{equation}
\sum_{\alpha=1}^\infty y_\alpha^i n_{k_\alpha^i}=q_i,\quad i=1,\ldots, r.
\label{diof}
\end{equation}
Notice that, once the $q_i$ are fixed, only a finite number of labels $k^i_\alpha$, $\alpha=1,\ldots,M_i(q_i):=M_i$, come into play in the equations (\ref{diof}). It may happen that some of these equations admit no solutions. In this case $2a=\sum_{i=1}^r q_i \sqrt{p_i}$ is such that $K(a)=0$ (i.e. $a$ does not belong to the area spectrum). On the other hand, if they do admit solutions, $2a=\sum_{i=1}^r q_i
\sqrt{p_i}$ is such that $K(a)>0$, the numbers
$k_\alpha^i$ tell us the allowed  values of the components of $\vec{k}$ involved, and the  $n_{k_\alpha^i}$ count the
number of components of $\vec{k}$ whose  values are equal to $k_\alpha^i$.

Let us now define the set
$$
\mathfrak{K}(a)=\BigTimes_{i=1}^r \mathfrak{K}_i(q_i)\,,\quad \mathfrak{K}_i(q_i)=\big\{\mathbf{k}_i=\{(k_\alpha^i,n_{k_\alpha^i})\}_{\alpha=1}^{M_i}\,:\,\sum_{\alpha=1}^{M_i} y_\alpha^i n_{k_\alpha^i}=q_i \big\}\,.
$$
Given $\mathbf{k}=(\mathbf{k}_1,\cdots,\mathbf{k}_r)\in\mathfrak{K}(a)$ it is possible to construct
$$
\frac{(\sum_{i=1}^r\sum_{(k,n_k)\in \mathbf{k}_i} n_k)!}{\prod_{i=1}^r\prod_{(k,n_k)\in \mathbf{k}_i} n_k!}\,
$$
sequences in $\mathcal{K}(a)$ by permuting the elements of the multiset\footnote{Here we are considering the sets $\mathbf{k}_i$ as a multisets. Hence $(k,n_k)\in \mathbf{k}_i$ should be understood as a notation signaling the presence of $n_k$  elements,  each of them equal to $k$, in the multiset.} $\cup_i \mathbf{k}_i$. Then, given $2a=\sum_{i=1}^r q_i \sqrt{p_i}$, the cardinality of $\mathcal{K}(a)$ is
$$
K(a)=|\mathcal{K}(a)|=\frac{(\sum_{i=1}^r\sum_{(k,n_k)\in \mathbf{k}_i} n_k)!}{\prod_{i=1}^r\prod_{(k,n_k)\in \mathbf{k}_i} n_k!}\,.
$$
In order to determine $N(a)$ from $K(a)$  it is enough  to realize that given $\vec{k}\in \mathcal{K}(a)$ each component $k_i$ gives rise to two different values $m_i\in\{-k_i/2,k_i/2\}$ of the corresponding components of $\vec{m}\in\pi^{-1}(\vec{k})$, whereas the number of allowed configurations once the projection constraint is incorporated can be easily obtained by using a simple generating function as in \cite{EF}. We have then the following

\bigskip

\begin{thm}
The value of $N(a)$ is given by
\begin{equation}
N(a)= \sum_{\mathbf{k}\in\mathfrak{K}(a)}\left(\frac{(\sum_{i=1}^r\sum_{(k,n_k)\in \mathbf{k}_i} n_k)!}{\prod_{i=1}^r\prod_{(k,n_k)\in \mathbf{k}_i} n_k!}\prod_{i=1}^r\prod_{(k,n_k)\in \mathbf{k}_i} 2^{n_k}\right)\,,
\label{N(a)}
\end{equation}
whereas, when the components of the sequences $\vec{m}$ are required to satisfy the projection constraint $\sum_i m_i=p$, we get
\begin{equation}
N(a,p):=[z^{2p}] \sum_{\mathbf{k}\in\mathfrak{K}(a)}\left(\frac{(\sum_{i=1}^r\sum_{(k,n_k)\in \mathbf{k}_i} n_k)!}{\prod_{i=1}^r\prod_{(k,n_k)\in \mathbf{k}_i} n_k!}\prod_{i=1}^r\prod_{(k,n_k)\in \mathbf{k}_i} (z^{k}+z^{-k})^{n_k}\right)\,.\\
\label{N(a,p)}
\end{equation}
\hfill$\square$
\end{thm}

\noindent We have used the notation $[z^{2p}]f(z)$ for the coefficient that multiplies $z^{2p}$ in the Laurent expansion of the function $f$. The numbers $N(a,p)$ can be conveniently encoded \cite{EF} in the generating function
\begin{eqnarray}
G(z;x_1,x_2,\dots)&=&\Big(\displaystyle 1-\sum_{i=1}^\infty\sum_{\alpha=1}^\infty (z^{k^i_\alpha} +z^{-k^i_\alpha}) x_i^{y^i_\alpha}\Big)^{-1}\,.\label{genN(a,p)}
\end{eqnarray}
The coefficient of the term $z^{2p}x_1^{q_1}\cdots x_r^{q_r}$ tells us the value of $N(a,p)$ for $2a=q_1\sqrt{p_1}+\cdots+q_r\sqrt{p_r}$. Moreover, using the relation $N(a)=\sum_{p\in \mathbb{Z}/2}N(a,p)$, the generating functions for the numbers $N(a)$ can be obtained from the generating function for the $N(a,p)$ numbers simply by setting the $z$-argument equal to one. Hence, the coefficient of the term $x_1^{q_1}\cdots x_r^{q_r}$ in the power expansion of
\begin{eqnarray}
G(1;x_1,x_2,\dots)&=&\Big(\displaystyle 1-2\sum_{i=1}^\infty\sum_{\alpha=1}^\infty  x_i^{y^i_\alpha}\Big)^{-1}\,\label{genN(a)}
\end{eqnarray}
gives us the value of $N(a)$ for $2a=q_1\sqrt{p_1}+\cdots+q_r\sqrt{p_r}$.
It is important to notice that $G(z;0,0,\dots)=1$. This can be interpreted as the extra one that appears in the prescription (\ref{entropylog}) of \cite{DL}. Formulas (\ref{N(a,p)}) and (\ref{N(a)}) summarize in a compact way the algorithms used in \cite{prlnos} to compute the black hole degeneracy spectrum. The same information is stored in the generating functions  (\ref{genN(a,p)}) and (\ref{genN(a)}) in a way that will let us write down very compact exact formulas for $N_{\leq}(a,p)$ and $N_{\leq}(a)$. This is the purpose of the next section.

\section{From $N(a), N(a,p)$ to $N_{\leq}(a),N_{\leq}(a,p)$}{\label{From}}

The coefficients of the power terms of the generating functions given above can be written in closed form as multiple contour integrals depending on $q_1,\ldots,q_r$ by using Cauchy's theorem. For example, given $2a=q_1\sqrt{p_1}+\cdots+q_r\sqrt{p_r}$ ,
\begin{eqnarray*}
N(a)&=&[x_1^{q_1}\cdots x_r^{q_r}]G(1;x_1,x_2,\ldots,x_r,0,\ldots)\\
&=&\frac{1}{(2\pi i)^r}\oint_{\gamma_1}\frac{\mathrm{d}\zeta_1}{\zeta_1^{q_1+1}}\,\cdots\oint_{\gamma_r}
\frac{\mathrm{d}\zeta_r}{\zeta_r^{q_r+1}}
G(1;\zeta_1,\zeta_2,\ldots,\zeta_r,0,\ldots)
\end{eqnarray*}
with appropriately chosen contours $\gamma_i$ surrounding the origin. Such integral representations are usually a good starting point to obtain asymptotic expansions in terms of the parameters appearing in them (the $q_i$ in this case). As we are really interested in the asymptotic behaviour of the entropy as a function of the area there is a necessary intermediate step: determining the parameters $q_i(a)$ as functions of the area. If these can be written in a reasonably simple closed form and are regular enough, just by plugging them in integral expressions of the type written above we could have closed formulas in terms of the area for the objects that we are interested in. The unfortunate fact is that the coefficients $q_i(a)$ as functions of the area oscillate wildly and in an rather unpredictable way so this direct approach is, to say the least, hard.

The sector $\mathcal{A}_{\mathrm{IH}}=\{a_n:n\in\mathbb{N}\}$ of the spectrum of the area operator relevant in the study of black hole entropy is a countable, \textit{ordered}  ($a_n<a_{n+1}$ for all $n\in \mathbb{N}$), subset of the real line. It is then possible, in principle, to build the sequence $\{N(a_n,p):n\in\mathbb{N}\}$. For a fixed value of the area spectrum $a_n$ we can then obtain $N_{\leq}(a_n,p)$ as
$$
N_{\leq}(a_n,p)=\sum_{i=1}^n N(a_i,p).
$$
In fact, if the values of $N(a_n,p)$ are encoded in the generating function $g_p(x)=\sum_{n\in \mathbb{N}}N(a_n,p)x^n$ this summation can be carried out by a well-known procedure consisting in taking the new generating function\footnote{For simplified models where the area spectrum is taken to be equally spaced this can be done \cite{Hanno}.}
$$
G_p(x)=\frac{g_p(x)}{1-x}=\sum_{n\in\mathbb{N}}N_{\leq}(a_n,p) x^n.
$$
Again this is difficult in the present case because to succeed with this approach one would need to have an appropriate way (i.e. a manageable closed formula) to find the numbers $q_i$ corresponding to the $n^{th}$ element of the ordered set of area eigenvalues $\mathcal{A}_{\mathrm{IH}}$. This can be seen to be equivalent to solving the following two problems:

\bigskip

i) Given an eigenvalue of the area $a\in \mathcal{A}_{\mathrm{IH}}$, how many smaller eigenvalues do exist? (we refer to this as the \textit{area ordering problem}).

\bigskip

ii) Given $A\in\mathbb{R}$, what are the values of the $q_i$ corresponding to the largest area eigenvalue $a\in\mathcal{A}_{\mathrm{IH}}$ satisfying $a\leq A$? (alternatively to the closest eigenvalue to $A$).

\bigskip

\noindent The last question, in particular, must be answered because in practice we want to treat the areas as a continuous real parameter. Although the problems just described are probably not unsurmountable, here we will avoid them and give a remarkably simple procedure to perform the required summations by using Laplace transforms\footnote{In the following $\mathcal{L}[F(a);s]$ denotes the Laplace transform, expressed in the variable $s$, of the function $F(a)$. On the other hand, $\mathcal{L}^{-1}[f(s);a]$ denotes the inverse Laplace transform of the function $f(s)$ in terms of the variable $a$.}. The key idea is to make use of the following two simple facts:

\bigskip

\noindent 1) $\displaystyle\mathcal{L}[\delta(a-\alpha);s]=e^{-\alpha s}$ for $\alpha\geq 0$. Here $\delta(a-\alpha)$ denotes the Dirac delta distribution supported at $\alpha$.

\bigskip

\noindent 2) If $\displaystyle\mathcal{L}[F(a);s]=f(s)$, then the Laplace transform of $\int_0^a F(a')\,\mathrm{d}a'$ is simply $s^{-1}f(s)$.

\bigskip

\noindent\textbf{Generating functions and Laplace transform:}  Let  us consider the ``generating function''\footnote{We will not worry about regularity or convergence issues here.}
\begin{equation}
G(x)=\sum_{n\in \mathbb{N}}\beta_n x^{\alpha_n}\label{gf}
\end{equation}
defined in terms of the sequences  $\mathcal{A}=\{\alpha_n\,:\,n\in\mathbb{N}\}$, $0\leq \alpha_n<\alpha_{n+1}$,  and $\mathcal{B}=\{\beta_n \,:\, n\in\mathbb{N}\}$. Notice that, in general, the $\alpha_n$ are not necessarily integers. If we take the distribution
$$
F(a)=\sum_{n\in \mathbb{N}} \mathbb{\beta}_n \delta(a-\alpha_n)
$$
we have that
$$
F_\leq(a)=\int_0^a F(a')\mathrm{d}a'=\sum_{\{n:\alpha_n\leq a\}}\beta_n \theta(a-\alpha_n)=
\sum_{\{n:\alpha_n\leq a\}}\beta_n\,,\quad a\not\in \mathcal{A}\,,
$$
represents  the sum of the numbers $\beta_n$ corresponding the the values of $\alpha_n$ smaller than $a$ (for $a\not\in \mathcal{A}$; notice that, otherwise, the integral in the above formula for $F_\leq$ is ill-defined). In order to extend the definition of $F_\leq$ to the values of $a\in\mathcal{A}$ as the sum $\sum_{\{n:\alpha_n\leq a\}}\beta_n$  it suffices to consider the limit from the right
$$
F_\leq(a)=\lim_{A\rightarrow a^+}\int_0^A F(a')\mathrm{d}a'=\,\sum_{\{n:\alpha_n\leq a\}}\beta_n\,,\quad a\geq0.
$$
If $\beta_n\geq0$ for every $n\in\mathbb{N}$ the existence of these limits is guaranteed by the fact that $F_{\leq}$ is an increasing function. If not all the values of $\beta_n$ are equal this function has jump singularities in the values $\alpha_n$. It is important to realize that the spacing between the values of $\alpha_n$ plays no role in the previous formula, so it works equally well for evenly or unevenly spaced values of $\alpha_n$. Also it is important to realize that under mild conditions on the sequences $\mathcal{A}$ and $\mathcal{B}$ the function $\sum_{\{n:\alpha_n< a\}}\beta_n \theta(a-\alpha_n)$ will be exponentially bounded and its Laplace transform well defined.

In order to compute $F_\leq(a)$ the idea is then to consider the function
$$
P(s)=G(e^{-s})=\sum_{n\in \mathbb{N}}\beta_n e^{-\alpha_n s},
$$
defined in terms of (\ref{gf}). Notice that it encodes the information about both sequences   $\mathcal{A}$ and $\mathcal{B}$. We can think about $P(s)$ as the Laplace transform
$$
\mathcal{L}[\sum_{n\in \mathbb{N}}\beta_n\delta(a-\alpha_n);s]=P(s).
$$
The arguments given above immediately tell us that
$$
F_{\leq}(a)=\mathcal{L}^{-1}[s^{-1}P(s);a]=\mathcal{L}^{-1}[s^{-1}\sum_{n\in \mathbb{N}}\beta_ne^{-\alpha_n s};a]
$$
if $a$ does not coincide with any of the values corresponding to the sequence $\mathcal{A}$. On the other hand if $a\in\mathcal{A}$ then the fact that at jump singularities the inverse Laplace transform always gives the average between the left and right limits means that $\mathcal{L}^{-1}[s^{-1}P(s);a]\neq\sum_{\{n:\alpha_n\leq a\}}\beta_n$. As mentioned above this can be simply fixed by taking appropriate limits from the right.
$$
F_{\leq}(a)=\lim_{A\rightarrow a^+}\mathcal{L}^{-1}[s^{-1}P(s);A]\,.
$$
A trivial but useful comment is that $F_\leq$ is constant in intervals that do not contain points of $\mathcal{A}$.

\bigskip

\noindent \textbf{Example:} A concrete example of the previous procedure is the following. Let us consider the sequences  $\mathcal{A}=\{\alpha_n=n-1\,:\, n\in \mathbb{N}\}$ and  $\mathcal{B}=\{\beta_n=1\,:\, n\in \mathbb{N}\}$ associated with the generating function
$$G(x)=\sum_{n\in \mathbb{N}} x^{n-1}=\frac{1}{1-x}\,.$$
In this case
$$
F_\leq(a)=\sum_{\{n\in \mathbb{N}\,:\, n-1\leq a\}} 1= \theta(a)(\lfloor a\rfloor +1)\,,\quad a\geq0.
$$
On the other hand
$$P(s)=G(e^{-s})=\sum_{n\in \mathbb{N}} e^{-(n-1)s}=\frac{1}{1-e^{-s}}$$
and it is clear that, for a non negative $a\not\in \mathbb{N}_0$,
\begin{eqnarray*}
\mathcal{L}^{-1}[s^{-1}P(s); a]&=&\frac{1}{2\pi i}\int_{x_0-i\infty}^{x_0+i \infty}\frac{e^{as}\, \mathrm{d}s}{s(1-e^{-s})}\quad (\mathrm{where}\,\, x_0>0)\\
&=&\sum_{ \{ k\in \mathbb{N}_0 \,:\, k\leq a\}}  \theta(a-k)=\theta(a)(\lfloor a\rfloor +1)\\
&=&\theta(a)\left(a+\frac{1}{2}+\frac{1}{\pi}\sum_{k=1}^\infty \frac{\sin 2\pi k a}{k}\right)\,,
\end{eqnarray*}
where the last equality is obtained by using residues to compute the integral in the first line. As
we can see in this case the inverse Laplace transform formula gives $a+1/2$ for integer values of $a$ (i.e. the average $(F_\leq(a+0)+F_\leq(a-0))/2$ of the left and right limits).\hfill $\blacksquare$

\bigskip

\noindent\textbf{Laplace transform and black hole entropy:} The scheme presented above can be used compute
$$
N_{\leq}(a)=\mathcal{L}^{-1}[s^{-1}\sum_{n\in \mathbb{N}}N(a_n)e^{-a_n s};a]\,,
$$
if $a$ does not coincide with any of the values corresponding to the spectrum of the area operator, and extend the previous formula to all the positive values of the area --according to the entropy definition that we have adopted-- just by taking limits from the right
$$
N_{\leq}(a_n)=\lim_{a\rightarrow a_n^+}\mathcal{L}^{-1}[s^{-1}\sum_{i\in \mathbb{N}}N(a_i)e^{-a_i s};a]\,,\quad a_n\in \mathcal{A}_{\mathrm{IH}}.
$$
Notice that for $N(a_n)\geq 0$ the previous limits are always well defined.

The key point now is to realize that by using our generating functions, in particular $G(1;x_1,x_2,\ldots)$, we can get a simple expression for $\sum_{n\in \mathbb{N}}N(a_n)e^{-a_n s}$.
To this end it is enough to substitute the arguments $x_i$ in $G(1;x_1,x_2,\ldots)$ by
$x_i=e^{-s\sqrt{p_i}/2}$. This is so because $x_1^{q_1}\cdots x_r^{q_r}\mapsto e^{-\frac{s}{2}(q_1\sqrt{p_1}+\cdots+q_r\sqrt{p_r})}=e^{-as}$ when $2a=q_1\sqrt{p_1}+\cdots+q_r\sqrt{p_r}$. By doing this we find
$$
P(s):=\sum_{n\in \mathbb{N}}N(a_n)e^{-a_n s}+1=G(1;e^{-s\sqrt{p_1}/2},e^{-s\sqrt{p_2}/2},\dots)=\Big(\displaystyle 1-2\sum_{i=1}^\infty\sum_{\alpha=1}^\infty e^{-sy^i_\alpha\sqrt{p_i}}\Big)^{-1}\,.
$$
The exponentials $e^{-sy^i_\alpha\sqrt{p_i}}$ appearing in this function can be simplified by taking into account that the numbers $(k_\alpha^i,y_\alpha^i)$ are solutions to the Pell equation and hence
$$
y^i_\alpha\sqrt{p_i}=\sqrt{k^i_\alpha(k^i_\alpha+2)}\,.
$$
This way ge get
\begin{eqnarray*}
P(s)=\Big(\displaystyle 1-2\sum_{i=1}^\infty\sum_{\alpha=1}^\infty  e^{-s\sqrt{k^i_\alpha(k^i_\alpha+2)}/2}\Big)^{-1}
=\Big(\displaystyle 1-2\sum_{k=1}^\infty e^{-s\sqrt{k(k+2)}/2}\Big)^{-1}\,,
\end{eqnarray*}
where we have used the fact that the values of the $k$'s appearing in the solutions to the Pell equations corresponding to different squarefree integers $p_i$ are always different (so that $\{k_\alpha^i:\alpha\in\mathbb{N}\}\cap\{k_\alpha^j:\alpha\in\mathbb{N}\}=\emptyset$ whenever $i\neq j$) and also that every $k\in\mathbb{N}$ appears in the solution to \textit{some} Pell equation because $\sqrt{k(k+2)}$ can always be written as the product of a positive integer times a SRSFN.

In order to take into account the projection constraint it is convenient to take $z=e^{i\omega/2}$, that in practice lets us get $N(a,p)$ by performing an integral around a contour in the complex $z$-plane consisting of a unit circumference surrounding the origin (notice that in this case $z=1$ can be obtained by choosing $w=0$). By doing this we get the function
\begin{eqnarray*}
P(s,\omega):=G(e^{i\omega/2};e^{-s\sqrt{p_1}/2},e^{-s\sqrt{p_2}/2},\dots)&=&\Big(\displaystyle 1-\sum_{i=1}^\infty\sum_{\alpha=1}^\infty (e^{i\omega k^i_\alpha/2} +e^{-i\omega k^i_\alpha/2}) e^{-sy^i_\alpha\sqrt{p_i}}\Big)^{-1}\,.
\end{eqnarray*}
The exponentials $e^{-sy^i_\alpha\sqrt{p_i}}$ appearing in this function can be simplified if we use the Pell equations as before so we get
$$
P(s,\omega)=\Big(\displaystyle 1-2\sum_{i=1}^\infty\sum_{\alpha=1}^\infty  e^{-s\sqrt{k^i_\alpha(k^i_\alpha+2)}/2}\cos \frac{\omega k^i_\alpha}{2}\Big)^{-1}\!\!=\!\!\Big(\displaystyle 1-2\sum_{k=1}^\infty  e^{-s\sqrt{k(k+2)}/2}\cos\frac{\omega k}{2}\Big)^{-1}\,.
$$
Notice that $P(s,0)=P(s)$. Finally, by performing the sums as explained above we have that
\begin{eqnarray}
s^{-1}P(s,\omega)
&=&s^{-1}\Big(\displaystyle 1-2\sum_{k=1}^\infty  e^{-s\sqrt{k(k+2)}/2}\cos\frac{\omega k}{2}\Big)^{-1}\nonumber\\
&=& P_\leq(s,\omega)+s^{-1}\,
\label{P(s,omega)}
\end{eqnarray}
gives us the Laplace-Fourier transform $P_\leq(s,\omega)$ of $N_\leq(a,p)$ plus a $s^{-1}$ extra term that originates in the additional $1$ appearing in the Domagala-Lewandowski prescription for $\mathfrak{n}(a)=N_\leq(a,0)+1$. By inverting these expressions we then get the following result

\begin{thm}{\label{th1}} When $a\not\in \mathcal{A}_{IH}$, the values of $N_{\leq}(a)$, $N_{\leq}(a,p)$, and $\mathfrak{n}(a)$ are given by
\begin{eqnarray*}
N_\leq(a)&=&\frac{1}{2\pi i}\int_{x_0-i\infty}^{x_0+i\infty} \Big(s^{-1}P(s,0)-s^{-1}\Big) e^{as}  \,\mathrm{d}s \\
&=&\frac{1}{2\pi i}\int_{x_0-i\infty}^{x_0+i\infty} s^{-1}\Big(\displaystyle 1-2\sum_{k=1}^\infty e^{-s\sqrt{k(k+2)}/2}\Big)^{-1} e^{sa}  \,\mathrm{d}s-\theta(a)\,.\\
N_\leq(a,p)&=&\frac{1}{8\pi^2 i}\int_{0}^{4\pi} \int_{x_0-i\infty}^{x_0+i\infty} \Big(s^{-1}P(s,\omega)-s^{-1}\Big) e^{as} e^{-ip\omega} \,\mathrm{d}s \,\mathrm{d}\omega\\
&=&\frac{1}{8\pi^2 i}\int_{0}^{4\pi} \int_{x_0-i\infty}^{x_0+i\infty}
s^{-1}\Big(\displaystyle 1-2\sum_{k=1}^\infty e^{-s\sqrt{k(k+2)}/2}\cos\frac{\omega k}{2}\Big)^{-1}e^{as} e^{-ip\omega}\,\mathrm{d}s \,\mathrm{d}\omega\noindent\\
&&-\delta(p,0)\theta(a)\,.\\
\mathfrak{n}(a)&=&\frac{1}{8\pi^2 i}
\int_{0}^{4\pi} \int_{x_0-i\infty}^{x_0+i\infty} s^{-1}P(s,\omega) e^{as}  \,\mathrm{d}s \,\mathrm{d}\omega\\
&=&\frac{1}{8\pi^2 i}\int_{0}^{4\pi} \int_{x_0-i\infty}^{x_0+i\infty}
s^{-1}\Big(\displaystyle 1-2\sum_{k=1}^\infty e^{-s\sqrt{k(k+2)}/2}\cos\frac{\omega k}{2}\Big)^{-1} e^{as}\,\mathrm{d}s \,\mathrm{d}\omega\,\,,\quad a\geq 0,
\end{eqnarray*}
\noindent where the value $x_0>0$ is chosen to the right of all the singularities of the integrand in order to guarantee that the previous integrals converge. On the other hand, for $a_n\in \mathcal{A}_{IH}$, the values $N_{\leq}(a_n)$, $N_{\leq}(a_n,p)$, and $\mathfrak{n}(a_n)$ coincide with the $\lim_{a\rightarrow a_n^+}$ of the above expressions. \hfill$\square$
\end{thm}

At this point we should compare these results with the ones obtained by Meissner \cite{M}. First of all we see that once the extra $1$ that appears in the definition of the entropy is incorporated the expression that we find for $N_{\leq}(a)$ \textit{exactly} coincides with the one given by him. However, the expression that we find in the case when the projection constraint is taken into account differs from the one that appears in \cite{M}. The difference between both expressions --once the extra $1$ is taken into account-- just amounts to an extra cosine factor in formula (35) of \cite{M} that should have been (using our $P_\leq$ for Meissner's $P$)
\begin{eqnarray*}
P_\leq(s,\omega)&=&\noindent \\
&&\hspace{-1.5cm}\frac{2}{s}\sum_{k=1}^\infty \left(\exp(-s\sqrt{k(k+2)/4})\cos\frac{\omega k}{2}\right)\left(1-2\sum_{k=1}^\infty\exp(-s\sqrt{k(k+2)/4})\cos\frac{\omega k}{2}\right)^{-1}\,.
\end{eqnarray*}
Though this is somewhat speculative, there are two possible sources for this discrepancy: a simple typographical error or, more likely, an artifact introduced by the approximations that, as Meissner himself acknowledges, have been used in \cite{M} to derive his formula (35). In our opinion the only way to settle this issue is to revisit Meissner's derivation and get his formulas again. This is the purpose of the next section.

\section{Functional equations for $N_\leq(a,p)$ and $N_\leq(a)$}{\label{functional}}

The main goal of this section is to obtain the results of the previous one by using functional equations as in \cite{M}. We start by first considering the obtention of $N_{\leq}(a)$. The key idea is to pick a value for $a$ and relate the values of $N_{\leq}(a)$ to those corresponding to $a^\prime<a$. Let us consider the sequences in $\mathcal{N}_{\leq}(a)$ and classify them according to their first element. This allows us to partition this set as a disjoint union $$\mathcal{N}_{\leq}(a)=\bigcup_{k\in\mathbb{N}}\mathcal{N}_{\leq}^{(k)}(a)$$
with
\begin{eqnarray*}
\mathcal{N}^{(k)}_\leq(a)&:=&\{\vec{m}\,|\,\exists n\in \mathbb{N}\,:\,  \vec{m}\in (\mathbb{Z}_*/2)^n\,,\, \sum_{i=1}^n\sqrt{|m_i|(|m_i|+1)}\leq a, |m_1|=k/2\}\,.
\end{eqnarray*}
Notice that $\mathcal{N}^{(k)}_\leq(a)=\emptyset$ and $N^{(k)}_\leq(a)=0$ if $a<\sqrt{3}/2$.
The cardinalities of $\mathcal{N}_{\leq}(a)$ and $\mathcal{N}^{(k)}_\leq(a)$ (denoted by $N^{(k)}_\leq(a)$) are then related by
$$
N_{\leq}(a)=\sum_{k=1}^\infty N^{(k)}_\leq(a)\,.
$$
Notice also that the previous union involves, in practice, only a finite number of nonempty $\mathcal{N}^{(k)}_\leq(a)$ sets\footnote{For the same reason the previous sum is finite.} because given a fixed real value of $a\geq \sqrt{3}/2$ the maximum of $k=2|m_1|$ for a sequence in $\mathcal{N}_\leq(a)$ is $k_{\mathrm{max}}:=\lfloor \sqrt{1+4a^2}-1\rfloor$.

For every value of $k\leq k_{\mathrm{max}}$, and as long as $a\geq\sqrt{3}/2$, there are two sequences in $\mathcal{N}^{(k)}_\leq(a)$ consisting of a single element, viz, $(\pm k/2)$. The rest of them, if they exist, are of the form $\vec{m}=(\pm k/2,m_2,\ldots)$ and have, at least, two elements. If for each of the latter one considers the finite sequence $\vec{m}_*=(m_2,\ldots)$ the condition that $\vec{m}\in \mathcal{N}^{(k)}_\leq(a)$ is equivalent to $\vec{m}_*\in \mathcal{N}_{\leq}(a-\sqrt{k(k+2)}/2)$ so we conclude that the cardinality of each of the $\mathcal{N}^{(k)}_\leq(a)$ is simply given by
\begin{equation}
N^{(k)}_\leq(a)=2+2N_{\leq}(a-\sqrt{k(k+2)}/2).
\label{Nsupk}
\end{equation}
At this point it is convenient to extend the definition of $N_{\leq}(a)$ to arbitrary real values of $a$ in such a way that $N_{\leq}(a)=0$ for $a\leq 0$. This condition can be conveniently encoded as\footnote{Remember that with our definition for the function $\theta$ we have $\theta(0)=1$.}
\begin{eqnarray}
N_\leq(a)&=&\theta(a-\sqrt{3}/2)N_\leq(a)\,,\label{N_2}
\end{eqnarray}
and allows (\ref{Nsupk}) to account for the case in which the sets $N^{(k)}_\leq(a)$ consist of sequences with a single element.

Adding up the values of $N^{(k)}_\leq(a)$ given by equation (\ref{Nsupk}) we get
\begin{equation}
N_{\leq}(a)=\sum_{k=1}^{k_{\mathrm{max}}}N^{(k)}_\leq(a)=2\lfloor\sqrt{1+4a^2}-1\rfloor+2\sum_{k=1}^{k_{\mathrm{max}}}
N_{\leq}(a-\sqrt{k(k+2)}/2)\,,\quad a\geq \sqrt{3}/2\,.
\label{recurr}
\end{equation}

Equation (\ref{recurr}) for the function $N_\leq:\mathbb{R}\rightarrow \mathbb{N}_0$ can be conveniently rewritten as
\begin{eqnarray}
N_\leq(a)&=& 2\lfloor\sqrt{4a^2+1} -1\rfloor\theta(a-\sqrt{3}/2)+2\sum_{k=1}^\infty N_\leq(a-\sqrt{k(k+2)}/2)\,,\label{recN(a)1}
\end{eqnarray}
where (\ref{N_2}) allows us to extend the sum to infinity and the $\theta(a-\sqrt{3}/2)$ factor is needed in the first term or the right hand side of (\ref{recN(a)1}) to guarantee that it is zero for arbitrary negative values of $a$. Several comments are in order now. First of all it must be pointed out that formulas (\ref{recN(a)1}) and (\ref{N_2}) correspond to equation (5) of Meissner \cite{M}. We have carefully avoided to include the factor $\theta(a-\sqrt{3}/2)$ in (\ref{recN(a)1}) as in \cite{M} because it will be very useful to do so when deriving similar functional equations for $N_\leq(a,p)$. Notice also that condition (\ref{N_2}) can be substituted by one of the type
$$
N_\leq(a)=\theta(a-a_0)N_\leq(a)
$$
with $0<a_0\leq\sqrt{3}/2$ that is basically equivalent to requiring that $N_\leq(a)=0$ for $a\leq0$.

We solve now the previous functional equations by using Laplace transforms. Although here we will just reproduce the correct equation (13) of Meissner, we give some details that will be relevant when discussing the resolution of the functional equations for $N_\leq(a,p)$. As already stated in \cite{M} the fact that  $N_\leq(a)$ is exponentially bounded \cite{DL} and piecewise continuous guarantees that its Laplace transform exists and is well defined in a half-plane $\{s\in\mathbb{C}:\mathrm{Re}(s)>x_0\}$ for some $x_0\in\mathbb{R}$. We have then

\begin{eqnarray}
P_\leq(s)&:=&\int_{[0,\infty)} N_\leq(a)e^{-as}\,\mathrm{d}a\label{P(s)}\\
&=&2\int_{[\frac{\sqrt{3}}{2},\infty)} \lfloor\sqrt{4a^2+1} -1\rfloor e^{-as}\,\mathrm{d}a+2\int_{[\frac{\sqrt{3}}{2},\infty)}\Big( \sum_{k=1}^\infty N_\leq(a-\sqrt{k(k+2)}/2)\Big)e^{-as}\, \mathrm{d}a\nonumber\\
&=& \frac{2}{s}\sum_{k=1}^\infty e^{-s\sqrt{k(k+2)}/2}+2 \sum_{k=1}^\infty\int_{[\frac{\sqrt{3}}{2},\infty)} N_\leq(a-\sqrt{k(k+2)}/2)\,e^{-as}\,\mathrm{d}a \nonumber
\\
&=&\frac{2}{s}\sum_{k=1}^\infty e^{-s\sqrt{k(k+2)}/2}+2 \sum_{k=1}^\infty e^{-s\sqrt{k(k+2)}/2}\int_{[-\sqrt{k(k+2)}/2,\infty)}\!\!\!N_\leq(a) e^{-as}\,\mathrm{d}a\nonumber\\
&=&\frac{2}{s}\sum_{k=1}^\infty e^{-s\sqrt{k(k+2)}/2}+2P_\leq(s)\sum_{k=1}^\infty e^{-s\sqrt{k(k+2)}/2}\nonumber\,.
\end{eqnarray}
Here we can change the order between sums and integrations in the two integrals appearing in the second line as a consequence of the Beppo Levi theorem\footnote{The Beppo Levi theorem is a corollary of the monotonous convergence theorem for Lebesgue integrals and states that if $\{f_n\}_{n\in\mathbb{N}}$ is a sequence of non-negative measurable functions then
$$
\sum_{n=1}^\infty \int f_n=\int \sum_{n=1}^\infty f_n
$$}. We have also used the fact that $N_\leq(a)=0$ for $a\leq0$ to set the lower limits in the integrals equal to zero in the last but one line of (\ref{P(s)}). In fact this is the only condition that we need to solve the functional equation (\ref{recN(a)1}). We finally get \cite{M}
\begin{equation}
P_\leq(s)=\frac{1}{s}\left(1-2\sum_{k=1}^\infty e^{-s\sqrt{k(k+2)}/2}\right)^{-1}-\frac{1}{s}=\frac{2\sum_{k=1}^\infty e^{-s\sqrt{k(k+2)}/2}}{s\left(1-2\sum_{k=1}^\infty e^{-s\sqrt{k(k+2)}/2}\right)}\,.
\label{PP(s)}
\end{equation}
The fact that $P_\leq(s)$ is a proper Laplace transform tells us that we can write
\begin{equation}
N_\leq(a)=\frac{1}{2\pi i}\lim_{A\rightarrow a^+}\int_{x_0-i\infty}^{x_0+i\infty}P_\leq(s)e^{As}\,\mathrm{d}s .
\label{Nleq(a)}
\end{equation}
for some $x_0\in \mathbb{R}$ chosen in such a way that the singularities in the integrand are to the left of the integration contour in (\ref{Nleq(a)}). As we can see we recover \textit{precisely} the same result obtained in the previous section by using our generating functions.

\bigskip

A functional equation for $N_\leq(a,p)$ can be obtained in a similar way. Again we classify the sequences in $\mathcal{N}_\leq(a,p)$ according to the first element and partition this set as the disjoint union
$$\mathcal{N}_{\leq}(a,p)=\bigcup_{k\in\mathbb{Z}_*}\mathcal{N}_{\leq}^{(k)}(a,p)$$
with
\begin{eqnarray*}
\mathcal{N}^{(k)}_\leq(a,p)&:=&\{\vec{m}\,|\,\exists n\in \mathbb{N}\,:\,  \vec{m}\in (\mathbb{Z}_*/2)^n\,,\, \sum_{i=1}^n\sqrt{|m_i|(|m_i|+1)}\leq a\,,
\sum_{i=1}^n m_i=p \,, m_1=k/2\}\,.
\end{eqnarray*}
These are empty sets if $a<\sqrt{3}/2$ and also if $a<\sqrt{|p|(|p|+1)}$. As before the cardinality of $\mathcal{N}_{\leq}(a,p)$ can be written in terms of  $N_{\leq}^{(k)}(a,p)$ as
$$
N_{\leq}(a,p)=\sum_{k\in\mathbb{Z}_*}N_{\leq}^{(k)}(a,p)\,,
$$
where the previous sum is, again, finite.

Let us suppose now that $p\neq0$ and $a\geq\sqrt{3}/2$. If $k=2p$ the only sequence of length one belonging to $\mathcal{N}_{\leq}^{(k)}(a,p)$ is $(p)$; furthermore this can only happen if $\sqrt{|p|(|p|+1)}\leq a$. If they exist, the remaining sequences in $\mathcal{N}_{\leq}^{(k)}(a,p)$ have, at least, two elements and are of the form $\vec{m}=(k/2,m_2,\ldots)$. We have now that $\vec{m}\in\mathcal{N}_{\leq}^{(k)}(a,p)$ if and only if $(m_2,\ldots)\in \mathcal{N}_{\leq}(a-\sqrt{k(k+2)}/2,p-k/2)$. If $k\neq 2p$ there are no one-element sequences and the rest of them have, again, the form $\vec{m}=(k/2,m_2,\ldots)$.  We can then write
\begin{eqnarray*}
N_{\leq}^{(k)}(a,p)&=&\delta(k,2p)\theta(a-\sqrt{|p|(|p|+1)})
+N_\leq(a-\sqrt{|k|(|k|+2)}/2,p-k/2)\,,\quad p\neq 0\,.
\end{eqnarray*}
Here, as before, it is useful to extend the definition of $N_\leq(a,p)$ to $\mathbb{R}$ and take this into account by imposing the condition $N_\leq(a,p)=0$ for $a\leq 0$.  The previous reasoning is essentially valid in the $p=0$ case, the only difference is that, as the elements of the sequences are non-zero half integers, it is impossible now to have unit length sequences. Summarizing we find that
\begin{eqnarray*}
N_{\leq}^{(k)}(a,p)&=&(1-\delta(0,p))\delta(k,2p)\theta(a-\sqrt{|p|(|p|+1)})
+N_\leq(a-\sqrt{|k|(|k|+2)}/2,p-k/2)\,.
\end{eqnarray*}
Adding up for all the possible values of $k$ we get
\begin{eqnarray}
N_{\leq}(a,p)&=&\sum_{k\in\mathbb{Z}_*}N_{\leq}^{(k)}(a,p)\nonumber\\
&=&(1-\delta(0,p))\theta(a-\sqrt{|p|(|p|+1)})
+\sum_{k\in\mathbb{Z}_*}N_\leq(a-\sqrt{|k|(|k|+2)}/2,p-k/2)\nonumber\\
&=&(1-\delta(0,p))\theta(a-\sqrt{|p|(|p|+1)})\label{func_p}\\
&+& \sum_{\ell=1}^\infty \left(N_\leq(a-\sqrt{\ell(\ell+2)}/2,p-\ell/2)+
N_\leq(a-\sqrt{\ell(\ell+2)}/2,p+\ell/2)\right)\,.\nonumber
\end{eqnarray}
Notice that it is immediate to check that by summing in $p\in\mathbb{Z}/2$ one recovers the functional equation (\ref{recN(a)1}) from (\ref{func_p}) because
$$
\sum_{p\in \mathbb{Z}/2}\Big(1-\delta(0,p)\Big)\theta(a-\sqrt{|p|(|p|+1)})=2\lfloor\sqrt{4a^2+1} -1\rfloor\theta(a-\sqrt{3}/2)\,.
$$

It may seem a little bit surprising that we are not including another condition explicitly stating that $N_{\leq}(a,p)=0$ if $a<\sqrt{|p|(|p|+1)}$ as done in \cite{M}. In fact it is easy to see that this is a consequence of the functional relation (\ref{func_p}) and the condition $N_\leq(a,p)=0$ for $a\leq 0$ just by repeatedly using it to compute $N_{\leq}(a,p)$ from values corresponding to smaller $p$'s. The fact that we do not need to include a prefactor in the recurrence relation (\ref{func_p}) is the reason why it is indeed possible to get an exact solution to this functional equation without having to use any approximation. In our opinion the statement appearing in \cite{M} claiming that one has to use approximations to solve the functional relation for $N_\leq(a,p)$ stems from the difficulties in dealing with that prefactor. Let us solve (\ref{func_p}) subject to the condition
$N_\leq(a,p)=0$ for $a\leq 0$. First notice that the fact that we can write
$$
N_\leq(a)=\sum_{p\in \mathbb{Z}/2} N_\leq(a,p)
$$
with $N_\leq(a)$ exponentially bounded implies that $\sum_{p\in \mathbb{Z}/2}e^{i\omega p}N_\leq(a,p)$
is also exponentially bounded because
$$
\Big|\sum_{p\in \mathbb{Z}/2}e^{i\omega p}N_\leq(a,p)\Big|\leq \sum_{p\in \mathbb{Z}/2}N_\leq(a,p)= N_\leq(a)\,,
$$
and, hence, the previous sum has a well defined Laplace transform. The resolution of (\ref{func_p}) is carried out as follows.
\begin{eqnarray}
\hspace*{-1cm}P_\leq(s,\omega)&:=&\int_{[0,\infty)} e^{-sa}\Big[\sum_{p\in \mathbb{Z}/2}e^{i\omega p}N_\leq(a,p)\Big]\mathrm{d}a\nonumber\\
&=& \int_{[0,\infty)} e^{-sa}\Big[\sum_{p\in \mathbb{Z}/2}e^{i\omega p}  (1-\delta(0,p))\theta(a-\sqrt{|p|(|p|+1)})\Big]\mathrm{d}a\nonumber\\
&+&\int_{[0,\infty)} e^{-sa}\Big[\sum_{p\in \mathbb{Z}/2}e^{i\omega p} \sum_{m\in\mathbb{Z}_*}N_\leq(a-\sqrt{|m|(|m|+2)}/2,p-\frac{m}{2})\Big]\mathrm{d}a\nonumber\\
&=& \frac{2}{s}\sum_{k=1}^\infty e^{-s\sqrt{k(k+2)}/2}\cos{\frac{k\omega}{2}}+\!\!\!\sum_{p\in \mathbb{Z}/2}\!\!e^{i\omega p}\!\!\int_{[0,\infty)} \!\!\!\!\!\!\!e^{-sa}\!\!\! \sum_{m\in\mathbb{Z}_*}N_\leq(a-\sqrt{|m|(|m|+2)}/2,p-\frac{m}{2})\mathrm{d}a\nonumber\\
&=& \frac{2}{s}\sum_{k=1}^\infty e^{-s\sqrt{k(k+2)}/2}\cos{\frac{k\omega}{2}}+\sum_{p\in \mathbb{Z}/2}e^{i\omega p}\sum_{m\in\mathbb{Z}_*} e^{-s\sqrt{|m|(|m|+2)}/2}
\int_{[0,\infty)}\!\!\!\! e^{-sa}N_\leq(a,p-\frac{m}{2})\mathrm{d}a\nonumber\\
&=& \frac{2}{s}\sum_{k=1}^\infty e^{-s\sqrt{k(k+2)}/2}\cos{\frac{k\omega}{2}}+2\sum_{k=1}^\infty e^{-s\sqrt{k(k+2)}/2}\cos \frac{k\omega}{2}\!\!\sum_{p^\prime\in \mathbb{Z}/2}\!\!
e^{i\omega p^\prime}\!\!\int_{[0,\infty)} \!\!e^{-sa}N_\leq(a,p^\prime)\mathrm{d}a\nonumber\\
&=& \frac{2}{s}\sum_{k=1}^\infty e^{-s\sqrt{k(k+2)}/2}\cos{\frac{k\omega}{2}}+2P_\leq(s,\omega)\sum_{k=1}^\infty e^{-s\sqrt{k(k+1)}/2}\cos \frac{k\omega}{2}\,.\nonumber
\end{eqnarray}
Here we can justify to change the order of sums and integrals as a consequence of the dominated convergence theorem. We have also used the fact that $N_\leq(a,p)=0$ for $a\leq0$ in order to write the lower limits in the previous integrals equal to zero as we did for $P_\leq(s)$. It is important to notice at this point that this is the only condition that we need to impose in order to solve the previous functional relations. In particular we have been able to do this \textit{without} the
$\theta(a-\sqrt{|p|(|p|+1)}$ prefactor used by Meissner. In our opinion this is crucial to avoid the use of simplifying assumptions. According to the previous derivation we get
\begin{equation}
P_\leq(s,\omega)=\frac{2}{s}\left(\sum_{k=1}^\infty e^{-s\sqrt{k(k+2)}/2}\cos{\frac{k\omega}{2}}\right)\left(1-2\sum_{k=1}^\infty e^{-s\sqrt{k(k+1)}/2}\cos \frac{k\omega}{2}\right)^{-1}\,,
\label{PP(s,omega)}
\end{equation}
in perfect agreement with the result obtained in the previous sections by using generating functions.
We can get $N_\leq(a,p)$ from (\ref{PP(s,omega)}) by inverting the Laplace-Fourier transform and recover the result of theorem \ref{th1}.

\section{Analytic properties of $P(s)$}{\label{poles}}

We study in this section the analytic structure of the function $P(s)$. This is a first necessary step to understand the behaviour of $P(s,\omega)$. We start by enunciating and proving several lemmas concerning the poles of $P(s)$ and finally give the main result of this section concerning the accumulation of the real parts of these poles precisely to the value $\tilde{\gamma}_M$. This result is important in the asymptotic analysis of the black hole entropy. In the following we will denote
$$
Q(s):=\frac{1}{P(s)}=1-2\sum_{k=1}^\infty e^{-s\sqrt{k(k+2)}/2}\,,\quad \mathrm{Re}(s)>0\,.
$$
This is obviously an analytic function in any band $\mathrm{Re}(s)\geq x_0>0$ due to the uniform convergence of the series $\sum_{k=1}^\infty e^{-s\sqrt{k(k+2)}/2}$. In fact
$$|e^{-s\sqrt{k(k+2)}/2}|=e^{-\textrm{Re}(s)\sqrt{k(k+2)}/2}\leq e^{-x_0\sqrt{k(k+2)}/2}$$
for all $s$ in the strip $\mathrm{Re}(s)\geq x_0>0$. Hence, using Weierstrass criterion, $\sum_{k=1}^\infty e^{-s\sqrt{k(k+2)}/2}$ converges uniformly in  $\mathrm{Re}(s)\geq x_0>0$ to an analytic function (see, for example, \cite{Gay} for a good review of convergence properties of Dirichlet series).  In fact, it can be proved that $Q$  is an analytic almost periodic function \cite{AP}. Remember that an analytic function $f(s)=f(x+iy)$, regular in a strip $x_1<x<x_2$ ($-\infty\leq x_1<x_2\leq +\infty$), is called \textit{almost periodic} (uniformly almost periodic) if for e
very $\varepsilon$ there exists a length $L=L(\varepsilon)$   such that every interval $y_0<y<y_0+L$ of length $L$ on the imaginary axis contains at least one translational number  $\tau=\tau(\varepsilon)$ associated with $\varepsilon$, i.e., a number $\tau$ satisfying the inequality
$$
|f(s+i\tau)-f(s)|<\varepsilon
$$
for all $s$ in the strip $x_1<x<x_2$. Every  periodic function, such as $f_k(s)=e^{-s\sqrt{k(k+2)}/2}$, is almost periodic. Also a uniformly convergent sequence of almost periodic functions, such as $\sum_{k=1}^\infty e^{-s\sqrt{k(k+2)}/2}$, is almost periodic \cite{AP}.

\bigskip

Prior to giving the main result of this section we state and prove some simple but useful lemmas.

\begin{lem}
Let $F$ be the restriction of $Q$ to the positive real axis $\mathbb{R}^+=(0,\infty)$, then $F$ is an analytic and monotonically growing function.
\end{lem}

\noindent This is so because $1-F$ is the limit of a sum of strictly monotonically decreasing functions of the type $e^{-x\sqrt{k(k+2)/2}}$.\hfill$\square$

\bigskip

\begin{lem}
$F$ has a single zero in $\mathbb{R}^+$ and, hence, $Q$ has only one real zero.
\end{lem}

\noindent This is an immediate consequence of the continuity and monotonicity  of $F$ and the fact that
$$
\lim_{x\rightarrow 0^+}F(x)=-\infty\,,\quad \lim_{x\rightarrow \infty}F(x)=1\,.
$$
$\hfill\square$

\bigskip

In the following we will denote this zero as $\tilde{\gamma}_M$. It obviously satisfies
\begin{equation}
\sum_{k=1}^\infty e^{-\tilde{\gamma}_M\sqrt{k(k+2)}/2}=\frac{1}{2}.\label{gamma1}
\end{equation}

\begin{lem}
The only zero of $Q$ with real part equal to $\tilde{\gamma}_M$ is $\tilde{\gamma}_M$ itself.
\end{lem}

\noindent Let us suppose that there exists $\tilde{s}_0=\tilde{\gamma}_M+i\tilde{y}_0\in\mathbb{C}$, with $\tilde{y}_0\neq 0$ such that
$$
1-2\sum_{k=1}^\infty e^{-\tilde{s}_0\sqrt{k(k+2)}/2}=0\,.
$$
Then the following conditions must hold
\begin{eqnarray}
&&\sum_{k=1}^\infty e^{-\tilde{\gamma}_M\sqrt{k(k+2)}/2}\cos(\tilde{y}_0\sqrt{k(k+2)}/2)=\frac{1}{2}\,,\label{c}\\ &&\sum_{k=1}^\infty e^{-\tilde{\gamma}_M\sqrt{k(k+2)}/2}\sin(\tilde{y}_0\sqrt{k(k+2)}/2)=0\,.\label{s}
\end{eqnarray}
However, from (\ref{gamma1}) we must have
$$
\sum_{k=1}^\infty e^{-\tilde{\gamma}_M\sqrt{k(k+2)}/2}=\sum_{k=1}^\infty e^{-\tilde{\gamma}_M\sqrt{k(k+2)}/2}\cos(\tilde{y}_0\sqrt{k(k+2)}/2)\,,
$$
which is impossible. In fact
$$
e^{-\tilde{\gamma}_M\sqrt{k(k+2)}/2}\geq  e^{-\tilde{\gamma}_M\sqrt{k(k+2)}/2}\cos(\tilde{y}_0\sqrt{k(k+2)}/2)\,,\quad \forall k\in \mathbb{N}
$$
and taking into account the fact that $\sqrt{k(k+2)}$ is irrational\footnote{If $\tilde{y}_0\sqrt{k(k+2)}/2\equiv 0 \mod 2\pi$ for some $k$ we can always find $k_0\in \mathbb{Z}$ such that $\tilde{y}_0\sqrt{k_0(k_0+2)}/2\not\equiv 0 \mod 2\pi$.} and $\tilde{y}_0\neq0$, there exists $k_0\in \mathbb{N}$ such that
$$
e^{-\tilde{\gamma}_M\sqrt{k_0(k_0+2)}/2} >  e^{-\tilde{\gamma}_M\sqrt{k_0(k_0+2)}/2}\cos(\tilde{y}_0\sqrt{k_0(k_0+2)}/2)\,.
$$
We then conclude that
$$
\sum_{k=1}^\infty e^{-\tilde{\gamma}_M\sqrt{k(k+2)}/2}>\sum_{k=1}^\infty e^{-\tilde{\gamma}_M\sqrt{k(k+2)}/2}\cos(\tilde{y}_0\sqrt{k(k+2)}/2)\,,
$$
and, hence, (\ref{c}) cannot be satisfied. We then conclude that $\tilde{s}_0$ cannot be a zero of $Q$.
$\hfill\square$
\bigskip

\begin{lem}
The real part of the zeroes of $Q$ different from $\tilde{\gamma}_M$ is strictly smaller than $\tilde{\gamma}_M$.
\end{lem}

\noindent Let us take $\tilde{s}_0=\tilde{x}_0+i\tilde{y}_0\in \mathbb{C}$, $\tilde{s}_0\neq \tilde{\gamma}_M$, such that $Q(\tilde{s}_0)=0$. Then it follows that
$$
\frac{1}{2}=\left|\sum_{k=1}^\infty e^{-\tilde{s}_0\sqrt{k(k+2)}/2}\right|\leq \sum_{k=1}^\infty e^{-\tilde{x}_0\sqrt{k(k+2)}/2}\,.
$$
Using now (\ref{gamma1}), we find that
\begin{equation}
\sum_{k=1}^\infty e^{-\tilde{\gamma}_M\sqrt{k(k+2)}/2}\leq \sum_{k=1}^\infty e^{-\tilde{x}_0\sqrt{k(k+2)}/2}\,.\label{menor}
\end{equation}
Now, taking into account that the function $f:(0,\infty)\rightarrow \mathbb{R}:x\mapsto\sum_{k=1}^\infty e^{-x\sqrt{k(k+2)}/2}$ is continuous and monotonically decreasing, we see that equation (\ref{menor}) implies that
$$
\tilde{x}_0\leq \tilde{\gamma}_M\,.
$$
As we have assumed that $\tilde{s}_0\neq \tilde{\gamma}_M$ we conclude that $\mathrm{Re}(\tilde{s}_0)=\tilde{x}_0<\tilde{\gamma}_M$. $\hfill\square$

\bigskip

We end this section with the following important result

\bigskip

\begin{thm}
The function $Q$ has an infinite number of zeros. The set of the real parts of the zeros of $Q$ has an accumulation point at $\tilde{\gamma}_M$.
\end{thm}

An immediate consequence of almost periodicity and uniform convergence in a strip is that, if the equation
$Q(s) = 0$ is solvable in the strip $\mathrm{Re}(s)\geq x_0>0$, then it will have infinitely many solutions and their imaginary parts will form a relatively dense set\footnote{A set $T$ of real numbers $\tau$ is called \textit{relatively dense} \cite{AP} if there are no arbitrary gaps among the numbers $\tau$ or, to be exact, if some length $L$ exists such that every interval $(a,a+L)\subset \mathbb{R}$ of this length contains at least one number $\tau$  of the set $T$. Roughly speaking, a relatively dense set can be described as one that is just as dense as an arithmetic progression $\{\alpha n\,:\, n\in \mathbb{Z}\}$, $\alpha>0$.}. This is a
well-known application of Rouch\'{e}'s Theorem. Here we basically repeat the reasoning appearing in \cite{bombieri}. Let $\tilde{\gamma}_M$ be the real zero of $Q(s)=0$ in the strip $\mathrm{Re}(s)\geq x_0>0$. Then there
exists an $ r_0> 0$ such that the circumference $C_{\tilde{\gamma}_M}(r_0) = \{s\in \mathbb{C} \,: \,|s-\tilde{\gamma}_M| = r_0\} $ is contained in the strip, encircles a single zero of $Q(s)$, and
$Q (s) \neq 0$ on the points of $C_{\tilde{\gamma}_M}(r_0)$. Take now
$$\varepsilon:=\min\{|Q(s)|\,:\,s\in C_{\tilde{\gamma}_M}(r_0)\}\,.$$
By uniform almost periodicity, there exists $L>0$ such that every interval of length $L$ contains $\tau\in \mathbb{R}$ such that $|Q (s + i\tau ) - Q (s)| < \varepsilon$ along $C_{\tilde{\gamma}_M}(r_0)$ and, hence,
$$
|Q (s + i\tau ) - Q (s)|< |Q(s)|\,,\quad \forall s\in C_{\tilde{\gamma}_M}(r_0)\,.
$$
By Rouch\'{e}'s Theorem, we deduce that $Q (s)$  and $ Q(s + i\tau)$ have the same
number of zeros inside the circle $C_{\tilde{\gamma}_M}(r_0)$. By repeating the argument for every $0<r<r_0$ and taking into account that all zeros of $Q$ (except $\tilde{\gamma}_M$) satisfy $\mathrm{Re}(s)<\tilde{\gamma}_M$, we see that we can find zeros of $Q$ with real parts smaller but as close to $\tilde{\gamma}_M$ as we wish. $\hfill\square$

\bigskip

As shown in the following example (which is a simple extension of the example given above and of the same type as those given by Meissner in \cite{M}) the accumulation of the real parts of the poles in the integrand of an inverse Laplace transform changes its asymptotic behavior relative to the one that one would expect by considering only the real pole\footnote{A small perturbation of the previous function would allow us to have accumulating real parts different from the value corresponding to the real pole in the integrand.}.

\bigskip

\noindent \textbf{Example:} Let us consider now the sequences  $\mathcal{A}=\{\alpha_n=n-1\,:\, n\in \mathbb{N}\}$ and  $\mathcal{B}=\{\beta_n=2^{n-1}\,;\, n\in \mathbb{N}\}$ associated with the generating function
$$G(x)=\sum_{n\in \mathbb{N}} (2x)^{n-1}=\frac{1}{1-2x}\,.$$
In this case
$$
F_\leq(a)=\sum_{\{n\in \mathbb{N}\,:\, n-1\leq a\}} 2^{n-1}= \theta(a)(2^{\lfloor a\rfloor +1}-1)\,.
$$
On the other hand
$$P(s):=G(e^{-s})=\sum_{n\in \mathbb{N}} 2^{n-1} e^{-(n-1)s}=\frac{1}{1-2e^{-s}}$$
and we have now for a non negative $a\not\in \mathbb{N}_0$,
\begin{eqnarray*}
\mathcal{L}^{-1}[s^{-1}P(s); a]&=&\frac{1}{2\pi i}\int_{x_0-i\infty}^{x_0+i \infty}\frac{e^{as}\, \mathrm{d}s}{s(1-2e^{-s})}\quad (\mathrm{where}\,\, x_0>0)\\
&=&\sum_{ \{ k\in \mathbb{N}_0 \,:\, k<a\}}  2^k\theta(a-k)=\theta(a)(2^{\lfloor a\rfloor +1}-1)\nonumber\\
&=&\theta(a)2^a\left(\frac{1}{\log 2}+\sum_{k=1}^\infty\frac{4k\pi\sin(2ak\pi)+(2\log2)\cos(2ak\pi)}
{4\pi^2k^2+\log^2 2}\right)-1\,,
\end{eqnarray*}
where the last equality can be obtained by using residues to compute the integral.
We can see in this case that the inverse Laplace transform formula gives $3\cdot 2^{(a-1)}-1$ for integer values of $a$ (as expected the average $(F_\leq(a+0)+F_\leq(a-0))/2$ of the left and right limits).
The important issue now is to realize that $2^a-2^{\lfloor a\rfloor}$ oscillates with an \textit{exponentially growing} amplitude and hence the values of $F_\leq(a)$ for $a\rightarrow\infty$ are not simply proportional to $2^a$.
\hfill $\blacksquare$

\bigskip

\section{Conclusions}{\label{conclusions}}

In this paper we have discussed several ways to \textit{exactly} compute the black hole entropy in loop quantum gravity \cite{DL}. In particular we have given a procedure based on generating functions that gives an independent way to derive the results appearing in the literature on this issue. In this way we have been able to detect and correct a mistake in the expression appearing in \cite{M} for the black hole entropy when the projection constraint is taken into account. Second we have shown that the number-theoretical methods introduced in \cite{prlnos,EF}, and successfully used to get precise numerical information about the entropy for small black holes, can be used in an efficient way to obtain exact formulas in the spirit of \cite{M}. In a sense, the two approaches are unified in this paper. We would like to point out, anyway, that the kind of detailed information provided by the number-theoretical methods of \cite{prlnos} is very difficult to extract form the expressions of the entropy as integral transforms. Finally, we have discussed the analytic properties of some of the functions that appear in the expressions of the black hole entropy. The most important result in this respect concerns the distribution of poles in the integrand of the inverse Laplace transform defining the entropy. From this analysis we have shown that the value of the Immirzi parameter given in the literature is, in a sense, correct but the asymptotic analysis of the entropy may display an interesting behaviour superimposed on the expected linear growth of the entropy as a function of the area. This suggests that the entropy structure found in \cite{val1,prlnos} for small black holes in numerical computations may actually survive for macroscopic areas. We want to add several comments:

In our opinion the paper by Meissner is usually misread and misunderstood. In particular, the exponential form of the entropy in terms of the area is often taken as some kind of ansatz introduced to approximately solve the functional equations giving the black hole entropy (with and without the projection constraint). It is very important to emphasize that by using Laplace transforms or Fourier-Laplace transforms it is possible to get exact expressions for the entropy. The exponential ansatz can be used to quickly show that the sought exponential growth of the entropy is somehow present but should not be taken as a rigorous derivation of the asymptotic behaviour of the entropy.

Despite the claims by the author of \cite{M} it is possible to find an exact closed expression for the black hole entropy also when the projection constraint is included. We have shown this in two complementary ways. By using the generating functions given in \cite{EF} and by solving  a functional equation along the lines suggested in \cite{M}. In the latter case we have corrected an error in the original functional equation and rewritten it in a way that facilitates its exact resolution. Actually the solution given in \cite{M} is very close to the right one.

We have studied in detail the analytic structure of the functions appearing in the integrand of the contour integral that gives the entropy (in the simplified setting where the projection constraint is not incorporated) to see if its asymptotic expansion can be readily obtained by looking at the poles of the integrand as claimed in \cite{M}. We have proved several important results in this respect (in some cases completing the claims of Meissner):

\begin{itemize}

    \item[i)] There is indeed an infinite number of poles in the integrand of (\ref{P(s)}).

    \item[ii)] They are confined to a band in the complex plane and their real parts are bounded from above by $\tilde{\gamma}_M$.

    \item[iii)] There is only a single pole of the integrand with real part equal to $\tilde{\gamma}_M$.

    \item[iv)] The real parts of the poles have an accumulation point precisely for the value $\tilde{\gamma}_M$ (and maybe others).

\end{itemize}

The last point is specially important as far as the asymptotic behaviour of the entropy is concerned because in a situation such as the one described here it may not be true that the asymptotic behaviour is given by the contributions of the pole with the largest real part, in fact this is illustrated by the simple example given in section \ref{poles}. Notice also that even if this is the case the fact that one has other poles with real parts arbitrarily close to $\tilde{\gamma}_M$ means that there may be corrections that are relevant for macroscopic but not infinite areas. It is clear, nonetheless, that the exponential behaviour given in \cite{M}, should play a significant role in the final asymptotic form of the entropy. In particular the value of the Immirzi parameter $\tilde{\gamma}_M$ is distinguished by the fact that is the maximum of the real parts of the poles. In this sense it controls the growth of the entropy despite the fact that its asymptotic behaviour may not be given by a simple exponential.

The reader may argue that, in the end, the methods based on the solution of functional equations for the entropy as introduced by Meissner are rather quick and efficient so there is no need to resort to the kind of detailed combinatorial analysis of \cite{prlnos,EF}. Our opinion is that, although they are indeed very clever ways of tackling this hard problem, the kind of detailed information provided by our number-theoretical approach is very useful as shown by the fact that they provide an independent way to check the results obtained so far. Taken at face value the expressions for the black hole entropy as inverse integral transforms give the entropy at all scales (and hence also in the microscopic regime where the interesting behaviour of the entropy has been found \cite{val1}). However, due to the subtly oscillatory nature of the integrands they cannot be practically used to obtain the entropy with good precision. Our combinatorial methods are much better in this respect and, in any case it is possible to check that they give the same values for the entropy that the integral expressions (whenever they can be numerically computed). Also, they allow to exactly characterize the area spectrum and the microscopic configurations corresponding to the allowed values of the area going far beyond the results obtained in \cite{M}.

The issue of getting the right asymptotic and the behaviour of the entropy for macroscopic scales, for which the integral expressions that we give here are a good starting point, is the last important problem that remains to be addressed to completely understand the behaviour of the black hole entropy in LQG. This will be our goal in the immediate future.

\begin{acknowledgments}
The authors want to thank I. Agull\'o, E. F. Borja, J. D\'{\i}az-Polo, L. Freidel,  J. Lewandowski, G. Mena Marug\'an,  H. Sahlmann, T. Thiemann, and very specially to K. Meissner  for their comments and encouragement.

\end{acknowledgments}

\end{document}